\documentclass[a4paper,12pt]{article}
\usepackage[english]{babel}
\usepackage{fullpage}
\usepackage{epsfig}
\usepackage{amsfonts}
\usepackage{amssymb}
\usepackage{amstex}
\usepackage{theorem}

\numberwithin{equation}{section}

\newtheorem{theorem}{Theorem}[section]
\newtheorem{corollary}[theorem]{Corollary}

\newtheorem{proposition}[theorem]{Proposition}
{  \theorembodyfont{\normalfont}
 }
{  \theorembodyfont{\normalfont}
 }

\title{Symmetry and resonance in periodic FPU chains}
\author{Bob Rink \\
Mathematics Institute, Utrecht University, \\
  PO Box 80.010, 3508 TA Utrecht, The Netherlands, \\
Telephone: 003130-2534557, Fax: 003130-2518394.}
\begin{document}
\maketitle 

\abstract{\noindent The symmetry and resonance properties of the
Fermi Pasta Ulam chain with periodic boundary conditions are exploited
to construct a near-identity transformation bringing this
Hamiltonian  system into a
particularly simple form. This `Birkhoff-Gustavson normal form' retains the
symmetries of the original system and we show that in most cases this
allows us to view 
the periodic FPU Hamiltonian as a perturbation of a nondegenerate Liouville
integrable Hamiltonian. According to the
KAM theorem this
proves the existence of many invariant tori on which 
motion is quasiperiodic. Experiments confirm this qualitative
behaviour. We note that one can not expect it in
lower-order resonant
Hamiltonian systems. So the FPU chain is an exception and its special
features are caused by a 
combination of special resonances and 
symmetries.}

\section{Introduction}

\noindent The $n$ particles FPU chain with periodic boundary
conditions is a model for point masses moving on a circle with
nonlinear forces acting between the nearest neighbours. It is in fact
the $n$ degrees of freedom Hamiltonian system 
on $\mathbb{R}^{2n}$ induced by the real-analytic Hamiltonian  

\begin{equation}\label{hamfpu}
H = \sum_{j \in \mathbb{Z}/_{n \mathbb{Z}}} \frac{1}{2} p_j^2 +
V(q_{j+1} - q_j) \ ,  
\end{equation}
in which $V: \mathbb{R} \to \mathbb{R}$ is a real-analytic potential
energy function of the form 
\begin{equation}
V(x) = \frac{1}{2!} x^2 + \frac{\alpha}{3!} x^3 + \frac{\beta}{4!} x^4
+ \ldots \ \ . 
\end{equation}
The $\alpha, \beta, \ldots$ are real parameters measuring the
nonlinearity in the forces between the particles in the chain. \\ 
\\
\noindent Numerically, the FPU system was first studied by E. Fermi,
J. Pasta and S. Ulam, see \cite{Alamos}. These authors used the chain
as a model for a string of which the elements interact
in a nonlinear way. They expected that in the
presence of small nonlinearities, the chain 
would show ergodic behaviour, meaning that almost all orbits densely
fill up an energy-level set of the Hamiltonian. Ergodicity would eventually
lead to an equal distribution of energy between the 
various Fourier modes of the system, a concept called
{\it thermalisation}.  FPU's nowadays famous numerical experiment was
intended to
investigate at what timescale thermalisation would take place. The
result was astonishing: it turned out that there was no sign of
thermalisation at all. Putting initially all the energy 
in one Fourier mode, they observed that this energy was shared by only
a few other modes, the remaining modes were hardly
excited. Additionally, within a not too long time the system returned
close to its initial state. On increasing the strength of the
nonlinearity, this 
recurrence occurred even earlier. Later computations, e.g. described
in \cite{equipartition}, confirmed that the same phenomena
can also be observed in very large periodic chains. Empirical evidence
was found that for small total energy, normal 
mode energies are hardly shared. Ergodic behaviour can only be
observed when 
the energy level passes a certain critical value. \\ 
\indent In
1965 an 
article of Zabuski and Kruskal appeared, cf. \cite{Kruskal}. These
authors considered the Korteweg-de Vries 
equation as a continuum limit of the FPU chain and numerically
found the first indications for the stable behaviour of solitary
waves, thereby suggesting an explanation for the striking data of the FPU
experiment. In 1967, Gardner, Greene, Kruskal and Miura
(\cite{gardner}) discovered 
infinitely many conserved quantities for the KdV equation, which should account for
the regular behaviour of its solutions. Reference \cite{Palais}
contains a good overview of these results. They are suggestive, but
do not provide a full explanation of FPU's observations as the
impact of the transition from 
a discrete to a continuous chain has never been analysed.\\
\indent  There is another, possibly correct explanation for the
quasiperiodic behaviour of the FPU system. It is based on the
Kolmogorov-Arnol'd-Moser (KAM) 
theorem (cf. \cite{arnold}) and different from the Zabuski-Kruskal argument, it should work
especially well for 
chains with a low number of particles. As is well-known
(cf. \cite{arnold}), the general solution of 
an $n$ degrees of freedom Liouville
integrable Hamiltonian system is constrained to move in an
$n$-dimensional torus
 and is not at all ergodic but periodic or quasiperiodic.
The KAM theorem states that most of the invariant
tori of a nondegenerate integrable system persist under small Hamiltonian
perturbations. Thus many authors,
starting with Izrailev and Chirikov in 
\cite{Chirikov}, have
stated that the KAM theorem explains the observations of the FPU
experiment. This reasoning seems plausible, but, as was clearly
pointed out by Ford in  
\cite{Ford}, it is still completely unclear why the FPU system should
be a
perturbation of such a nondegenerate integrable system. This gap in the theory was recently mentioned again in the book of Weissert (\cite{Weissert}). \\ 
\indent What does `nondegenerate' mean here? Let us consider the frequency map
$\omega$, which 
assigns to each $n$-dimensional invariant torus of a Liouville
integrable system 
the $n$-dimensional vector of 
frequencies of the (quasi)periodic motion on this torus. An integrable
system is called 
`nondegenerate' if $\omega$ is a local diffeomorphism. The KAM theorem
holds for perturbations of these 
nondegenerate 
integrable systems. \\
\indent But it is no exception for an integrable system to be
degenerate. A common example is the harmonic oscillator of
which the frequency map 
is constant: the harmonic
oscillator is highly degenerate. And
indeed, perturbations of it are known 
that are ergodic even on low-energy level sets of the
Hamiltonian. Ford gives a nice example of such a perturbation
in his review article \cite{Ford}. We conclude that, although the FPU
Hamiltonian can 
be considered as
a perturbation of an integrable system -namely the harmonic
oscillator-, the KAM theorem does not apply here! \\
\indent The aim of this paper is to overcome this problem. The method
we use to do so is called {\it Birkhoff-Gustavson
normalisation} - it is sometimes also called {\it resonant normalisation}.
It provides a transformation of phase space that in many cases enables 
us to write 
the periodic FPU Hamiltonian as a perturbation of a nondegenerate integrable
Hamiltonian. \\
\indent  It must be stressed that it seems highly exceptional that one
can do this for a resonant Hamiltonian system such as the periodic FPU
chain. The current paper intends to make clear that the special
symmetry,  
eigenvalue and resonance characteristics of the periodic FPU system play a 
crucial role in 
the construction of the near-identity transformation. It turns out
that {\it these 
characteristics} 
cause the nondegenerate near-integrability of the chain. The
conclusion is that the KAM theorem applies 
because of 
these resonance and symmetry properties: the quasiperiodic behaviour
that Fermi, Pasta and Ulam observed is in some sense an
exceptional feature of the FPU system.

\subsection{Outline of the paper}
This paper is a continuation of \cite{Rink} in which normal forms of
small chains are computed and the KAM theorem is verified. We
generalize and explain the results of \cite{Rink} in this paper.\\
\indent In sections \ref{phonons}-\ref{normformendiscrsymm} the
necessary theory is formulated. We
start with an 
investigation of the eigenvalues (section \ref{phonons}) and the discrete
symmetries (section \ref{discrsym}) of the periodic FPU
chain. The concept of a Birkhoff-Gustavson normal form as an
approximation of a Hamiltonian system is explained in section
\ref{normalisation}. It will be shown that normal forms for the
periodic FPU chain exist that inherit its symmetry properties.   \\
\indent In the appendix, which is based on notes of Beukers,
number theory is used to compute all lower order
resonances in the eigenvalues. We exploit this in sections \ref{simdiag} and
\ref{restriction} to prove theorem \ref{eenvoud}, which forms the core of
this paper: it gives the restrictions that the
Birkhoff-Gustavson normal 
form of any Hamiltonian with the same eigenvalues and symmetries as the
periodic FPU chain, is subject to.\\
\indent These restrictions on the normal
form allow us to point out many 
near-integrals of the chain in section
\ref{integrals}. We finish with an analysis of the $\beta$-chain,
which is proved to be near-integrable in
section \ref{betaketen}. The
KAM nondegeneracy condition can 
easily be checked 
when the $\beta$-chain contains an odd number of particles. Some open
questions are formulated for the even $\beta$-chain.

\subsection{Acknowledgement}
\noindent The author thanks Frits Beukers, Richard
Cushman, Hans Duistermaat, Reinout Quispel, Theo 
Tuwankotta and Ferdinand Verhulst for useful comments and
discussions.

\section{Phonons}\label{phonons}
To establish the sign conventions that we shall stick to during our
analysis, some basic definitions follow here. For further reading on
Hamiltonian systems and a thorough explanation of these concepts, the
reader is referred to \cite{A&M}.\\
\indent We shall be considering Hamiltonian systems of
differential 
equations on 
$\mathbb{R}^{2n}$, the elements of which are denoted by $(q,
p) = (q_1,  \ldots, q_n, p_1, \ldots, p_n)$. On $\mathbb{R}^{2n}$ the
symplectic form
$\sigma := \sum_{j=1}^n dq_j \wedge dp_j$ is defined. Endowed with this
symplectic form $\mathbb{R}^{2n}$ is a symplectic space. Any
Hamiltonian function $H:\mathbb{R}^{2n} \to \mathbb{R}$ induces a
Hamiltonian vector field $X_H$ on $\mathbb{R}^{2n}$ which is defined
by $\sigma(X_H, \cdot) = dH$. Furthermore, for any two Hamiltonians
$F$ and $G$ the Poisson brackets are defined as $\{F, G\}:= \sigma(X_F,
X_G) = dF \cdot X_G = - dG \cdot X_F$. \\
\indent Keeping these definitions in mind, we now start our analysis of the
periodic FPU chain:\\
\\
\noindent In order to facilitate the equations of
motion induced by the periodic FPU Hamiltonian (\ref{hamfpu}), we apply a well-known
Fourier transformation $(q, p) \mapsto (\bar q, \bar
p)$. For $1 \leq j < \frac{n}{2}$ define  
\begin{align} \nonumber & \bar q_j = \sqrt{\frac{2}{n}} \sum_{k=1}^n
\cos(\frac{2jk\pi}{n}) q_k \ , \ \ \bar p_j = \sqrt{\frac{2}{n}}
\sum_{k=1}^n \cos(\frac{2jk\pi}{n}) p_k \ , \\ 
& \bar q_{n-j} = \sqrt{\frac{2}{n}} \sum_{k=1}^n
\sin(\frac{2jk\pi}{n}) q_k \ , \ \ \bar p_{n-j} = \sqrt{\frac{2}{n}}
\sum_{k=1}^n \sin(\frac{2jk\pi}{n}) p_k \ . 
\end{align}
Furthermore, define 
\begin{equation}
\bar q_n = \frac{1}{\sqrt{n}} \sum_{k=1}^n q_k \ , \ \  \bar p_n =
\frac{1}{\sqrt{n}} \sum_{k=1}^n p_k \ , 
\end{equation}
and if $n$ is even,
\begin{equation}
\bar q_{\frac{n}{2}} = \frac{1}{\sqrt{n}} \sum_{k=1}^n (-1)^k q_k \ ,
\ \ \bar p_{\frac{n}{2}} = \frac{1}{\sqrt{n}} \sum_{k=1}^n (-1)^k p_k \
. 
\end{equation}
The new coordinates $(\bar q, \bar p)$ are known as {\it `phonons'}. The
transformation to phonons is symplectic, that is $\sigma =
\sum_{j=1}^n d \bar q_j \wedge d \bar p_j$. For a proof,
cf. \cite{poggiruffo} or \cite{Rink}. In
phonon-coordinates, the Hamiltonian reads  
\begin{equation}\label{haminphonon}
H = 
\sum_{j=1}^{n} \frac{1}{2} (\bar p_j^2 + \omega_j^2 \bar q_j^2) +
H_3(\bar q_1, \ldots, \bar q_{n-1}) + H_4(\bar q_1, \ldots, \bar
q_{n-1}) + \ldots\ , 
\end{equation}
in which $H_k\ (k=2,3, \ldots)$ denotes the $k$-th order part of $H$;
for $j=1, \ldots, 
n$, the numbers $\omega_j$ are the eigenvalues of the linear periodic FPU
problem:  
\begin{equation}
\omega_j := 2 \sin(\frac{j \pi}{n}) \ .
\end{equation}
Exact expressions for $H_3$ and $H_4$ in terms of the $\bar q_j$ can
be found in the literature, cf. \cite{poggiruffo}. We do not repeat
them. \\
\indent The linearised equations  are the equations induced by
$H_2$. They read: 
\begin{equation}
\bar q_j\ ' = \bar p_j \ , \bar p_j\ ' = - \omega_j^2 \bar q_j \ .
\end{equation}
The $\bar q_j, \bar p_j\ (1 \leq j \leq n-1)$ are harmonics with frequency
$\omega_j$; $\bar p_n$ is constant, whereas $\bar q_n$ increases with
constant speed -note that $\omega_n = 0$. In fact, the linearised
equations are Liouville 
integrable, with integrals $E_j := \frac{1}{2} (\bar p_j ^2 +
\omega_j^2 \bar q_j^2)$. The nonlinear equations
($\alpha$ or $\beta$ unequal to zero) are much harder to analyse. The
$E_j$ are for instance no longer constants of motion.

\section{Reduction of a continuous symmetry group}
From (\ref{haminphonon}) we see that $H$ is independent of $\bar
q_n$, even if $\alpha, \beta, \ldots \neq 0$. This implies that $ \bar p_n$
is an integral of $H$.  
The set $\bar p_n^{-1}(\{0\})$ defines a $2n-1$ dimensional hyperplane in
$\mathbb{R}^{2n}$, invariant under the flow of both $X_H$ and $X_{\bar
p_n}$. The flow of $X_{\bar
p_n} =
 \frac{\partial}{\partial \bar q_n}$ induces a symplectic
$\mathbb{R}$-action on this hyperplane. The time-$t$ flow $e^{tX_{\bar
p_n}}$ is actually given by 

\begin{equation}
e^{tX_{\bar
p_n}}: \sum_{j=1}^n \left( \bar  q_j \frac{\partial}{\partial \bar  q_j}
+ \bar p_j
\frac{\partial}{\partial \bar p_j} \right) \mapsto \sum_{j=1}^{n}
\left(\bar q_j
  \frac{\partial}{\partial \bar q_j} + \bar p_j
\frac{\partial}{\partial \bar p_j}\right)
+ t \frac{\partial}{\partial \bar q_n} 
 \ , 
\end{equation}
or written out in the original coordinates:

\begin{equation}
e^{tX_{\frac{1}{\sqrt{n}}\sum p_k}}
: \sum_{j=1}^n \left( q_j \frac{\partial}{\partial q_j} + p_j
\frac{\partial}{\partial p_j} \right) \mapsto \sum_{j=1}^{n-1} \left( (q_j
+ \frac{t}{\sqrt{n}}) \frac{\partial}{\partial q_j} + p_j
\frac{\partial}{\partial p_j} 
\right) \ . 
\end{equation}

\noindent The orbits of this flow are the
lines $(\bar q, \bar p) + \mathbb{R} \frac{\partial}{\partial \bar
q_n}$. It is clear that the $2n-2$ dimensional hyperplane $\bar
q_n^{-1}(\{0\}) \cap \bar p_n^{-1}(\{0\}) \cong
\mathbb{R}^{2n-2}$ is transversal to these
orbits. Therefore, $\mathbb{R}^{2n-2}$ is a model for the space $\bar
p_n^{-1}(\{0\}) 
/_{\mathbb{R}}$ of $X_{\bar p_n}$-orbits lying in $\bar p_n^{-1}(\{0\})$. $
\mathbb{R}^{2n-2}$ inherits the symplectic structure $\tilde \sigma :=
\sum_{j=1}^{n-1} d \bar q_j \wedge d \bar p_j$ from
$\mathbb{R}^{2n}$. And since 
the FPU Hamiltonian $H$ is constant on the orbits of the flow of
$X_{\bar p_n}$, $H$ reduces to a Hamiltonian on $\mathbb{R}^{2n-2}$ given by 
\begin{equation}\label{gereduceerdeham}
H = 
\sum_{j=1}^{n-1} \frac{1}{2} (\bar p_j^2 + \omega_j^2 \bar q_j^2) +
H_3(\bar q_1, \ldots, \bar q_{n-1}) + H_4(\bar q_1, \ldots, \bar
q_{n-1}) + \ldots\ . 
\end{equation}
\noindent The reduced Hamiltonian (\ref{gereduceerdeham}) represents
the periodic FPU system from which the centre of mass motion has been
eliminated. \\
\indent Since $\omega_j^2 > 0 \
(1 \leq j \leq n-1)$, we conclude with the Morse-Lemma
(cf. \cite{A&M}) that the level sets of $H$ are $2n-3$ dimensional
spheres around the 
origin of  $\mathbb{R}^{2n-2}$. And since $H$ is a constant of motion
for the flow of $X_H$,
we see that the origin is a stable stationary point for the
reduced system 
induced by the reduced Hamiltonian (\ref{gereduceerdeham}).

\section{Discrete symmetries}\label{discrsym}
Apart from the continuous family of symmetries of the previous
section, the FPU Hamiltonian has some discrete symmetries. These have important
dynamical consequences. \\  
\indent The first discrete symmetry is a rotation symmetry. Let $T:
\mathbb{R}^{2n} \to \mathbb{R}^{2n}$ denote the circle permutation, the
unique linear mapping defined by  
\begin{equation}\label{Tactie}
T: \frac{\partial}{\partial q_j} \mapsto \frac{\partial}{\partial
q_{j-1}} \ , \ \  
\frac{\partial}{\partial p_j} \mapsto \frac{\partial}{\partial
p_{j-1}} \ . 
\end{equation}
$T$ is symplectic: $T^*\sigma = \sigma$. Furthermore, note that $T$
leaves $H$ invariant: $T^*H := H \circ T = H$. This 
implies that the Hamiltonian vector field $X_H$ induced by $H$ is
equivariant under $T: \ DT \cdot X_H = X_H \circ T$. In other words:
if $\gamma:\mathbb{R} \to \mathbb{R}^{2n}$ is an integral curve of
$X_H$, then $T \circ \gamma: \mathbb{R} \to \mathbb{R}^{2n}$ is an
integral curve of $X_H$. This is why we call $T$ a symmetry of
$H$. The same thing holds for the powers of $T$. The group $\langle T
\rangle := \{\mbox{Id}, T, T^2, \ldots, T^{n-1}\} \cong
\mathbb{Z}/_{n\mathbb{Z}}$ is a discrete symmetry group of $H$. \\ 
\indent We can point out another discrete symmetry, namely the
reflection $S: \mathbb{R}^{2n} \to \mathbb{R}^{2n}$ which is the unique
linear mapping sending 
\begin{equation} \label{Sactie}
S: \frac{\partial}{\partial q_j} \mapsto - \frac{\partial}{\partial
q_{n-j}} \ , \ \   
\frac{\partial}{\partial p_j} \mapsto - \frac{\partial}{\partial
p_{n-j}} \ . 
\end{equation}
$S$ is again a symplectic symmetry: $S^* \sigma = \sigma$ and 
$S^*H = H$. The group $\langle S 
\rangle := $ $ \{\mbox{Id},S\} $ $\cong $ $ \mathbb{Z}/_{2
\mathbb{Z}}$, whereas 
the full discrete symmetry group $\langle T,S \rangle$ $ :=$ $\{\mbox{Id},$
$T, T^2,$ $ \ldots, T^{n-1},$ $S, ST^2,$ $\ldots, ST^{n-1}\}$ $ \ \cong \
D_n$ is 
called the `$n$-th dihedral group'; its group structure is determined
by the relation $ST = T^{n-1}S$. The vector field $X_H$ is equivariant
under the elements of $\langle T, S \rangle$, that is $\langle T, S
\rangle$ maps integral curves of $X_H$ to integral curves of
$X_H$. \\
\\
\noindent The reader should note that $T$ and $S$ leave $\bar
q_n^{-1}(\{0\}) \cap \bar p_n^{-1}(\{0\})$ invariant. Therefore, $T$
and $S$ reduce to linear symplectic mappings on $\mathbb{R}^{2n-2}$ that
leave the reduced Hamiltonian invariant\footnote{\noindent 
The FPU Hamiltonian also has a reversing symmetry,
namely the mapping 
$R: \mathbb{R}^{2n} \to \mathbb{R}^{2n}$ given by
\begin{equation}
R: \frac{\partial}{\partial q_j} \mapsto \frac{\partial}{\partial q_j}
\ , \frac{\partial}{\partial p_j} \mapsto -\frac{\partial}{\partial
p_j} \ . 
\end{equation}
$R$ leaves the FPU Hamiltonian invariant, i.e. $R^*H = H$. $R$ is
anti-symplectic in the sense 
that $R^*\sigma = -\sigma$. This implies that the vector field $X_H$
is anti-equivariant under $R$: $DR \cdot X_H = - X_H \circ R$. In
other words: if $\gamma:\mathbb{R} \to \mathbb{R}^{2n}$ is an integral
curve of $X_H$, then $R \circ \gamma \circ (-\mbox{Id}): \mathbb{R}
\to \mathbb{R}^{2n}$ is an integral curve of $X_H$.
Since $R$ leaves $\bar
q_n^{-1}(\{0\}) \cap \bar p_n^{-1}(\{0\})$ invariant, $R$ reduces to an
anti-symplectic mapping on $\mathbb{R}^{2n-2}$ leaving the reduced
Hamiltonian invariant.  More information
on reversing symmetries can be found in \cite{Lamb}.}.

\section{Normalisation}\label{normalisation}
We shall study the reduced FPU system (\ref{gereduceerdeham}) using
Birkhoff-Gustavson 
normalisation. In fact, we shall construct a near-identity
transformation of phase-space allowing us to write the FPU Hamiltonian
in `normal form', meaning that it can be seen as a
perturbation of a rather simple system. The study of the truncated
normal form
-that is this simpler system- leads to important conclusions for the
original FPU system. For
instance, the solutions of the truncated normal form are
approximations of low-energetic solutions of the 
original system valid on a long time-scale. Integrals of the truncated
normal form are near-integrals of the
original system: on orbits of low energy, they are almost conserved
for a long time.  See \cite{verhulst} for an explanation and explicit
statements. Furthermore, 
the truncated normal form can help us
understanding bifurcation phenomena. And last but not least, if the
truncated normal form of the FPU chain is integrable in a
nondegenerate way, then the FPU chain is a perturbation of a
nondegenerate integrable system.  We may
apply the
KAM theorem then and conclude that almost all low-energetic solutions of
(\ref{gereduceerdeham}) are quasiperiodic and move on
tori. Conclusions of this type were drawn for the first time in
\cite{Rink}. \\
\\
The setting of normalisation is the following:\\ 
\indent Let $P_k$ be the set of all homogeneous $k$-th degree
polynomials in $(\bar q_1, \ldots,$ $\bar q_{n-1}, \bar p_1,$ $ \ldots,
\bar p_{n-1})$. The set of all power series without linear part, $P :=
\bigoplus _{k \geq 2} P_k$, is a Lie-algebra with the Poisson
bracket. For each $h \in P$ the adjoint representation $\mbox{ad}_h :
P \to P$ is the linear operator defined by $\mbox{ad}_h (H) = \{h,
H\}$. Note that whenever $h \in P_k$, then $\mbox{ad}_{h}: P_l \to
P_{k+l-2}$. \\ 
\indent The flow $e^{t X_h}$ of a Hamiltonian vector field $X_h$
induced by $h \in P - P_2$ is a symplectic near-identity
transformation in $\mathbb{R}^{2n-2}$. For its
action on an arbitrary Hamiltonian $H \in P$ we have  $\frac{d}{dt}
(e^{t X_h})^*H = dH \cdot X_h = -\mbox{ad}_h (H)$. This is a linear
differential equation in $P$ of which the solution is $(e^{t X_h})^*H
= e^{- t \mbox{ad}_h} H$. In particular the near-identity
`Lie-transformation' $e^{-X_h} = \mbox{Id} - X_h + \ldots $ transforms
$H$ into 
\begin{equation}\label{lietrafo}
H' := (e^{-X_h})^*H = e^{\mbox{ad}_h} H = H + \{h, H\} +
\frac{1}{2}\{h, \{h, H\}\} + \ldots \  \ . 
\end{equation}
Let us denote the $k$-th order part of the Hamiltonian $H$ -that is
the projection of $H$ on $P_k$- by $H_k$. If for instance $h \in P_3$,
then we obtain the formula $H'_k = $ $\sum_{m=0}^{k-2}
\frac{1}{m!}$ $(\mbox{ad}_h)^{m}$ $ (H_{k-m})$. We just gathered all terms
of equal degree in formula (\ref{lietrafo}).\\ 
\\
\noindent Assume now, as is the case for the reduced FPU Hamiltonian,
that $\mbox{ad}_{H_2}: P_k \to P_k$ is semisimple
(i.e. complex-diagonalisable) for every $k \geq 2$. Then $P_k =
\mbox{ker ad}_{H_2} \oplus \mbox{im ad}_{H_2}$. In particular $H_3$ is
uniquely decomposed as $H_3 = f_3 + g_3$, with $f_3 \in \mbox{ker
ad}_{H_2}, \ g_3 \in \mbox{im ad}_{H_2}$. Now choose a $h_3 \in P_3$
such that $\mbox{ad}_{H_2} (h_3) = g_3$. One could for example choose
$h_3 = \tilde g_3 := (\mbox{ad}_{H_2}|_{\mbox{im
ad}_{H_2}})^{-1}(g_3)$. But clearly the choice $h_3 = \tilde g_3 +
p_3$ suffices for any $p_3 \in \mbox{ker ad}_{H_2} \cap P_3$. For the
new Hamiltonian $H'$ we calculate from (\ref{lietrafo}) that $H'_2 =
H_2, \ H'_3 = f_3 \in \mbox{ker ad}_{H_2}, \ H'_4 = H_4 + \{h_3, H_3 -
\frac{1}{2} g_3\}$, etc. But now we can again write $H'_4 = f_4 + g_4$
with $f_4 \in \mbox{ker ad}_{H_2}, \ g_4 \in \mbox{im ad}_{H_2}$ and
it is clear that by a suitable choice of $h_4 \in P_4$ the
Lie-transformation $e^{-X_{h_4}}$ transforms our $H'$ into $H''$ for
which $H_2'' = H_2, \ H''_3 = f_3 \in \mbox{ker ad}_{H_2}$ and $H''_4
= f_4 \in \mbox{ker ad}_{H_2}$. Continuing in this way, we can for any
finite $r \geq 3$ find a sequence of symplectic near-identity
transformations $e^{-X_{h_3}}, \ldots, e^{-X_{h_r}}$ with the property
that $e^{-X_{h_k}}$ only changes the $H_l$ with $l \geq k$, whereas
the composition $e^{-X_{h_r}} \circ \ldots \circ e^{-X_{h_3}}$
transforms $H$ into $\overline{H}$ with the property that
$\overline{H}_k$ Poisson commutes with $H_2$ for every $2 \leq k \leq
r$. $\overline{H}$ is called a normal form of $H$ of order $r$. Its
study can give us useful information on low-energetic solutions of the
original Hamiltonian $H$. More on normalisation by Lie-transformations
can be found in \cite{churchillkummerrod}. 

\section{Normal forms and discrete symmetry} \label{normformendiscrsymm}
In section \ref{discrsym} we investigated the discrete symmetries of
the periodic FPU Hamiltonian. We saw that they 
reduce to symmetries of the reduced FPU system
on $\mathbb{R}^{2n-2}$. In this section we
show how one can construct normal forms of the reduced FPU Hamiltonian that
have the same symmetry properties as the reduced FPU Hamiltonian
itself. This will help us formulating some important restrictions on
the normal form in section \ref{restriction}. \\ 
\indent The symmetry properties are captured in the definition of
the symmetric subspace of $P$: 
\begin{align} \nonumber
&P^{ST} := \{ f \in P | \ S^*f = f , \ T^*f = f  \}\ .
\end{align}

\noindent  Note that the FPU
Hamiltonian is in $P^{ST}$. \\
\indent The next observation is that $S^*$ and
$T^*$ are Lie-algebra automorphisms of $P$: 
\begin{equation}\label{automorfisme}
S^*\{f,g\} = \{S^*f, S^*g\} \ ,
\ \ T^*\{f,g\} = \{T^*f, T^*g\} \ .
\end{equation}
simply because $T$ and $S$ are symplectic. Now take $f \in P^{ST}$ and $g \in P^{ST}$. Then
from (\ref{automorfisme}) it follows that $S^*\{f, g\} = \{S^*f,
S^*g\} = \{f, g\}$ and $T^*\{f, g\} = \{T^*f,
T^*g\} = \{f, g\}$. This means that $P^{ST}$ is a
Lie-subalgebra of $P$: if $f, g \in P^{ST}$, then $\{f, g\} \in
P^{ST}$. Alternatively stated: if $h \in P^{ST}$, then $\mbox{ad}_h:
P^{ST} \to P^{ST}$. In particular, $e^{\mbox{ad}_h}: P^{ST} \to
P^{ST}$.\\ 
\\
\noindent Since $\mbox{ad}_{H_2}$ leaves $P^{ST}$ invariant, we know
that 
$P^{ST} = ( \mbox{ker ad}_{H_2} \cap P^{ST} ) \oplus ( \mbox{im
ad}_{H_2} \cap P^{ST} )$. So if we decompose the third order part of
the FPU Hamiltonian as 
$H_3 = f_3 + g_3$ with $f_3 \in \mbox{ker ad}_{H_2}, \ g_3 \in
\mbox{im ad}_{H_2}$, then $f_3, \ g_3 \in P_3^{ST}$
automatically. $h_3 = \tilde g_3 = (\mbox{ad}_{H_2} | _{\mbox{im ad}
H_2})^{-1}(g_3)$ is the unique element of $\mbox{im ad}_{H_2} \cap
P_3^{ST}$ for which $\mbox{ad}_{H_2}(h_3) = g_3$. But since $\tilde
g_3 \in P_3^{ST}$, we find that $H' = (e^{-X_{\tilde g_3}})^*H =
e^{\mbox{ad}_{\tilde g_3}} H \in P^{ST}$. Of course the choice $h_3 =
\tilde g_3 + p_3$ also suffices for any $p_3 \in \mbox{ker ad}_{H_2}
\cap P_3^{ST}$. \\
\indent It should be clear that continuing this procedure, we can
produce normal forms $\overline{H} \in P^{ST}$ of $H$ up to any finite
order\footnote{Although the bookkeeping is a bit harder, one can
extend the previous argument to prove that the normal forms
can also be chosen invariant under $R^*$. For a complete proof,
cf. \cite{churchillkummerrod}.}.\\

\section{Simultaneous diagonalisation}\label{simdiag}
From (\ref{automorfisme}) we infer that
\begin{equation}
(T^* \circ \mbox{ad}_{H_2})(f) = T^*\{H_2, f\} = \{ T^*H_2, T^*f\} =
\{H_2, T^* f\} = (\mbox{ad}_{H_2} \circ T^*)(f) \ .
\end{equation}
So $\mbox{ad}_{H_2}$ and $T^*$
commute on $P_k$. Therefore $\mbox{ad}_{H_2}$ leaves the eigenspaces
of $T^*$ invariant and we can diagonalise $\mbox{ad}_{H_2}$ and $T^*$
simultaneously. This allows us to calculate the subspace $P_k \cap
\mbox{ker ad}_{H_2} \cap \mbox{ker}(T^* - \mbox{Id}) \subset P_k$ in
which $\overline{H}_k$ is contained and helps us formulate
some important restrictions on the normal form of the FPU Hamiltonian. \\
\indent In order to perform this simultaneous diagonalisation, we
introduce the {\it `superphonons'} $(z, \zeta)$. For $1 \leq j <
\frac{n}{2}$, define:  

\begin{align} \nonumber \label{superphonons}
&z_j := \frac{1}{2}(\bar p_j - i \bar p_{n-j})  + \frac{i
\omega_j}{2} ( \bar q_j - i \bar q_{n-j}) = \frac{1}{\sqrt{2n}} 
\sum_{k=1}^n e^{-\frac{2\pi i j k}{n}}(p_k +i \omega_j q_k) \\ 
&\zeta_j :=  \frac{1}{2 i \omega_j} (\bar p_j + i \bar p_{n-j})
-\frac{1}{2}(\bar q_j +i\bar q_{n-j}) = \frac{1}{i \omega_j \sqrt{2n}} 
\sum_{k=1}^n e^{\frac{2\pi i j k}{n}}(p_k -i \omega_j q_k) \\\nonumber
&z_{n-j} := -\frac{1}{2}(\bar p_j  -i \bar p_{n-j}) + \frac{i \omega_j}{2} (\bar
q_j - i \bar q_{n-j}) = - \frac{1}{\sqrt{2n}} 
\sum_{k=1}^n e^{-\frac{2\pi i j k}{n}}(p_k -i \omega_j q_k) \\   
&\zeta_{n-j} := \frac{1}{2 i \omega_j}( \bar p_j + i \bar p_{n-j}) +
\frac{1}{2}( \bar q_j+ i \bar q_{n-j}) = \frac{1}{i \omega_j\sqrt{2n}} 
\sum_{k=1}^n e^{\frac{2\pi i j k}{n}}(p_k +i \omega_j q_k) \nonumber
\end{align}

\noindent and if $n$ is even:

\begin{align}
&z_{\frac{n}{2}} := \frac{1}{\sqrt{2} i \omega_{\frac{n}{2}}} (\bar
p_{\frac{n}{2}}+ i 
\omega_{\frac{n}{2}} \bar q_{\frac{n}{2}} ) = \frac{1}{i
\omega_{\frac{n}{2}}\sqrt{2n}} 
\sum_{k=1}^n (-1)^k(p_k + i \omega_{\frac{n}{2}} q_k)   \\
&\zeta_{\frac{n}{2}} :=  \frac{1}{\sqrt{2}}(\bar
p_{\frac{n}{2}} - i
\omega_{\frac{n}{2}}\bar q_{\frac{n}{2}}) =
\frac{1}{\sqrt{2n}} 
\sum_{k=1}^n (-1)^k(p_k - i \omega_{\frac{n}{2}} q_k) \nonumber
\end{align}

\noindent One checks that $\{z_j,z_k \} = \{\zeta_j,\zeta_k \} =0$
and $\{z_j,\zeta_k\} = \delta_{jk}$, the Kronecker delta. So our
superphonons define canonical coordinates, i.e. $\tilde \sigma =
\sum_{j=1}^{n-1} z_j \wedge \zeta_j$. \\ 
\\
\noindent From (\ref{Tactie}) we infer that $T^*q_j = q_{j+1}$ and
$T^*p_j = p_{j+1}$, where $q_j, p_j: \mathbb{R}^{2n} \to \mathbb{R}$
are the coordinate functions. So from (\ref{superphonons}) we see that
\begin{align}\nonumber
T^*: \ \  z_j \mapsto  e^{\frac{2\pi i j}{n}} z_j,\ \ 
\zeta_j \mapsto  e^{-\frac{2\pi i j}{n}} \zeta_j, \ & \  z_{n-j} \mapsto
e^{\frac{2\pi i j}{n}} 
z_{n-j},\ \  \zeta_{n-j} \mapsto e^{-\frac{2\pi i j}{n}} \zeta_{n-j}\ , \\
z_{\frac{n}{2}} \mapsto - z_{\frac{n}{2}} \  \mbox{and} & \
\zeta_{\frac{n}{2}} \mapsto
-\zeta_{\frac{n}{2}} \ .
\end{align}

\noindent We conclude that $T^*$ acts diagonally on $(z,
\zeta)$-coordinates. And it acts diagonally on monomials in $(z,
\zeta)$: if $\Theta, \theta \in \{0, 1, 2, \ldots\}^{n-1}$ are
multi-indices, then
\begin{equation}
T^*: \ z^{\Theta} \zeta^{\theta} \mapsto e^{\frac{2 \pi i \mu(\Theta,
\theta)}{n}} \ 
z^{\Theta} \zeta^{\theta} \ , 
\end{equation}

\noindent $\mu$ being defined as:

\begin{equation}\label{mu}
\mu(\Theta,\theta) := \sum_{1 \leq j < \frac{n}{2}} j (\Theta_j +
\Theta_{n-j} - \theta_j - \theta_{n-j} ) + \frac{n}{2} (
\Theta_{\frac{n}{2}} - \theta_{\frac{n}{2}} )  \ \ \mbox{mod} \ n \ \
. 
\end{equation}

\noindent On the other hand one calculates:

\begin{equation}
H_2 = \sum_{1 \leq j < \frac{n}{2}} i \omega_j ( z_j \zeta_j - z_{n-j}
\zeta_{n-j} ) + i\omega_{\frac{n}{2}} z_{\frac{n}{2}}
\zeta_{\frac{n}{2}} \ . 
\end{equation}

\noindent So we also diagonalised $\mbox{ad}_{H_2}$ with respect to monomials:

\begin{equation}
\mbox{ad}_{H_2} :  \ z^{\Theta} \zeta^{\theta} \mapsto \nu(\Theta,
\theta) z^{\Theta} \zeta^{\theta} \ , 
\end{equation}
 
\noindent in which $\nu$ is defined as

\begin{equation}\label{nu}
\nu(\Theta, \theta) := \sum_{1 \leq j < \frac{n}{2}} i\omega_j (
\theta_j - \theta_{n-j} - \Theta_j + \Theta_{n-j} ) \ +
i \omega_{\frac{n}{2}} ( \theta_{\frac{n}{2}} - \Theta_{\frac{n}{2}} ) \
. 
\end{equation}

\noindent Monomials $z^{\Theta}\zeta^{\theta}$ commuting with $H_2$
-the ones for which  $\nu(\Theta, \theta) = 0$- are called {\it
resonant} monomials. They are particularly important because they
cannot be normalised away.\\  
\\

\section{Restrictions for symmetric normal forms}\label{restriction}
\label{eigenwaardenonderzoekengevolgen} 

\noindent From section \ref{normformendiscrsymm} we know that we can
transform the periodic FPU Hamiltonian into a discrete symmetric
normal form of any desired order. Suppose we did so up  
to order $r$. Then $\overline{H}_k \in P_k \cap \mbox{ker
ad}_{H_2} \cap \mbox{ker}(T^* - \mbox{Id})$ for any $2 \leq k \leq r$.
But since both $T^*$ and
$\mbox{ad}_{H_2}$ act diagonally in $(z,\zeta)$-coordinates, we know
that this $\overline{H}_k$ must be a linear combination of monomials
$z^{\Theta} \zeta^{\theta}$ for which  
\begin{equation}\label{munu}
|\Theta| + |\theta| = k \ , \ \ \mu(\Theta, \theta) = 0\ \mbox{mod} \
 n \ \ \mbox{and} \ \ \nu(\Theta,\theta) = 0 \ . 
\end{equation}

\noindent Extra restrictions on $\overline{H}_k$, with which we shall
deal later, arise 
from the fact that $\overline{H}_k$ 
can be chosen in the even smaller set $P^{ST}$ \footnote{and invariant
under $R^*$}. But first
we investigate which $\Theta$ and $\theta$ satisfy
(\ref{munu}). Because the $\omega_j$ in (\ref{nu}) are of the form $2i
 \sin (\frac{j \pi}{n})$, this is actually a number-theoretical
question that we shall solve for $|\Theta| + |\theta| =
2,\ 3,\ 4$. \\ 
\\
\noindent The quadratic case - i.e. $|\Theta| + |\theta| = 2$ - is
easy: since all the $\omega_j$ are different, we find from
$\nu(\Theta, \theta) = 0$ that the Lie-subalgebra $P_2 \ \cap \
\mbox{ker ad}_{H_2} \subset P_2$ is spanned by the monomials 
\begin{equation}\label{baselt}
z_j\zeta_j, \  z_{n-j} \zeta_{n-j}, \ z_j z_{n-j}, \ \zeta_{j}
\zeta_{n-j} \ (1 \leq j < \frac{n}{2}) \ \mbox{and} \  z_{\frac{n}{2}}
\zeta_{\frac{n}{2}}\ .
\end{equation}
\noindent $T^*$ acts diagonally on these basis-elements as follows:
\begin{align} \label{Topbas}
T^*: z_j \zeta_j \mapsto z_j \zeta_j  , \ z_{n-j} \zeta_{n-j} &\mapsto
z_{n-j} 
\zeta_{n-j}  , \ z_{\frac{n}{2}} \zeta_{\frac{n}{2}} \mapsto
z_{\frac{n}{2}} \zeta_{\frac{n}{2}} \ , \\ \nonumber
z_j z_{n-j} \mapsto e^{\frac{4 \pi i j}{n}}z_j
z_{n-j} & , \ \zeta_j 
\zeta_{n-j} \mapsto e^{-\frac{4 \pi i j}{n}}
\zeta_j \zeta_{n-j} \ .
\end{align}

\noindent The Lie-subalgebra $P_2 \cap \mbox{ker ad}_{H_2} \cap
\mbox{ker}(T^* - \mbox{Id}) = \mbox{span}\{z_j \zeta_j, z_{n-j}
\zeta_{n-j},z_{\frac{n}{2}} \zeta_{\frac{n}{2}}  \}$ is abelian. \\
\indent From
(\ref{Sactie}) and  (\ref{superphonons}) we 
calculate the action of $S^*$ on the coordinate-functions:
\begin{equation}
S^*: z_j \mapsto - i \omega_j \zeta_{n-j} , \ \zeta_j \mapsto
\frac{1}{i \omega_j} z_{n-j} , \ z_{n-j} \mapsto  i \omega_j \zeta_j
, \ \zeta_{n-j} \mapsto \frac{-1}{i \omega_j} z_j , \ z_{\frac{n}{2}}
\mapsto - z_{\frac{n}{2}}, \ \zeta_{\frac{n}{2}} \mapsto -
\zeta_{\frac{n}{2}} \ .
\end{equation}
\noindent So the action on the basis-elements reads:
\begin{align}
S^*: z_j \zeta_j \mapsto - z_{n-j} \zeta_{n-j}  , \ z_{n-j}
\zeta_{n-j} &\mapsto 
- z_{j} 
\zeta_{j}  , \ z_{\frac{n}{2}} \zeta_{\frac{n}{2}} \mapsto
z_{\frac{n}{2}} \zeta_{\frac{n}{2}} \ , \\ \nonumber
z_j z_{n-j} \mapsto \omega_j^2 \zeta_j \zeta_{n-j} & , \ \zeta_j 
\zeta_{n-j} \mapsto \frac{1}{\omega_j^2} z_j z_{n-j} \ .
\end{align} 
\noindent We conclude that the Lie-subalgebra $P_2^{ST} \cap \mbox{ker
ad}_{H_2}$ is spanned by the quadratics $z_j \zeta_j - z_{n-j}
\zeta_{n-j}$ and $z_{\frac{n}{2}} \zeta_{\frac{n}{2}}$. Note that
$H_2$ itself is indeed a linear combination of 
these quadratics. \\
\\
\noindent The analysis is harder if we consider the cases $|\Theta| +
|\theta| = 3, \ 4$. With the use of number theory, the proof of the
following theorem is given in the appendix.  
\begin{theorem}\label{hoofdstellingeigenwaarden} \hspace{1cm}\\
\noindent i) The set of multi-indices $(\Theta, \theta) \in
\{0,1,2,\ldots\}^{2n-2}$ for which $|\Theta| + |\theta| = 3, \ \mu(\Theta,
\theta) = 0\ \mbox{mod}\ n$ and $\nu(\Theta, \theta) = 0$ is empty.\\
\\
\noindent ii) The set of
multi-indices $(\Theta, \theta) \in \{0,1,2,\ldots\}^{2n-2}$ for which
$|\Theta| + |\theta| = 4, \ \mu(\Theta, \theta) = 0\ \mbox{mod} \ n$ and
$\nu(\Theta, \theta) = 0$ is contained in the set given by the
relations $\theta_j - \theta_{n-j} - \Theta_j + \Theta_{n-j} =
\theta_{\frac{n}{2}} - \Theta_{\frac{n}{2}} = 0$. 
\end{theorem}
Theorem \ref{hoofdstellingeigenwaarden} has some major
implications. We shall investigate these now and they will be
summarised in theorem \ref{eenvoud}.
\\
\\
\noindent From i) we see that $P_3^{ST} \cap \mbox{ker
ad}_{H_2} \subset P_3 \cap \mbox{ker ad}_{H_2} \cap \mbox{ker}(T^* -
\mbox{Id}) = \{0\}$.\\
\indent First of all, this implies that we can always
transform away 
$H_3$ from the periodic FPU Hamiltonian: $\overline{H}_3 = 0$. This is an
unexpected result. Consider for example the chain with $6$ particles,
which satisfies a third order resonance relation:
$\omega_1:\omega_3:\omega_5 = 1:2:1$. For systems with a third order
resonance relation one can generally not expect $\overline{H}_3$ to be
trivial. But, as was observed for the first time in $\cite{Rink}$, it
{\it is trivial} for the $6$ particles chain. One could say that the
$1:2:1$-resonance is not active at $H_3$-level. {\it We now know that for
the periodic FPU chain no
resonance will ever be active at $H_3$-level.} This
simplification is caused by the 
symmetries of the FPU system. \\
\indent Secondly, we conclude from
i) that the $h_3$ of section
\ref{normformendiscrsymm} is uniquely determined by the requirement
that it be in $P_3^{ST}$. This in turn
uniquely determines 
$\overline{H}_4$. \\
\\
\noindent From ii) we infer that any element of $P_4
\cap \mbox{ker ad}_{H_2} \cap \mbox{ker}(T^* - \mbox{Id})$ must be a
linear combination of products of two of the basis-elements in
(\ref{baselt}). \\
\indent  Note
however that not all these products are really $T^*$-invariant and
that the full normal form is even invariant under $S^*$. We
work out these extra restrictions now. \\
\indent  The question which products of the basis-elements
(\ref{baselt}) are invariant under $T^*$ is easy to answer with help
of the formulas (\ref{Topbas}). Clearly, all products of $z_j \zeta_j,\
z_{n-j} \zeta_{n-j}$ and $z_{\frac{n}{2}} \zeta_{\frac{n}{2}}$
are. $T^*$ multiplies the terms $(z_j
\zeta_j)(z_k z_{n-k})$, $(z_j
\zeta_j)(\zeta_k \zeta_{n-k})$, $(z_{n-j}
\zeta_{n-j})(z_k z_{n-k})$, $(z_{n-j}
\zeta_{n-j})(\zeta_k \zeta_{n-k})$, $(z_{\frac{n}{2}}
\zeta_{\frac{n}{2}})(z_k z_{n-k})$ and $(z_{\frac{n}{2}}
\zeta_{\frac{n}{2}})(\zeta_k \zeta_{n-k})$ with a factor $e^{\pm \frac{4 \pi i
k}{n}} \neq 1$, so these terms are not invariant under $T^*$. $T^*$
multiplies $(z_j z_{n-j})(\zeta_k \zeta_{n-k})$ by $e^{\frac{4 \pi i
(j-k)}{n}}$ which is $1$ if and only if $2(j-k)=0$ mod $n$. But
because  $1 \leq j,k<\frac{n}{2}$, the condition is $2(j-k) = 0$, i.e.
$j=k$. Thus we end up with a term that we already had: $(z_j
z_{n-j})(\zeta_j \zeta_{n-j}) = (z_j \zeta_j)(z_{n-j}\zeta_{n-j})$. Finally,
the terms $(z_j z_{n-j})(z_k z_{n-k})$ and $(\zeta_j
\zeta_{n-j})(\zeta_k \zeta_{n-k})$ are multiplied by a factor $e^{\pm
\frac{4\pi i (j+k)}{n}}$ which is $1$ if and only if $2(j+k) = 0$ mod
$n$. But since $1 \leq j , k < \frac{n}{2}$, the only possibility is
that $2(j+k) = n$, that is $n$ must be even and
$j+k=\frac{n}{2}$. This concludes our search for fourth order
monomials invariant under $T^*$ and Poisson commuting with $H_2$.\\
\indent  We shall check now which
combinations of these terms are also invariant under $S^*$. The action
of $S^*$ on $P_2\cap \mbox{ker ad}_{H_2}$ can be diagonalised in real
coordinates. For this purpose, besides our familiar complex basis,
we also define the 
following real basis-elements for $P_2 \ \cap \ \mbox{ker 
ad}_{H_2}$. For $1 \leq j < \frac{n}{2}$, let 
\begin{align}\label{abcd}\nonumber
&a_j := i(z_j \zeta_j - z_{n-j} \zeta_{n-j}) = \frac{1}{2 \omega_j}
(\bar p_j^2 + \bar p_{n-j}^2 + \omega_j^2 \bar q_j^2 + \omega_j^2 \bar
q_{n-j}^2) \ , \\ 
&b_j := i(z_j \zeta_j + z_{n-j} \zeta_{n-j}) = \bar p_j \bar
q_{n-j} - \bar p_{n-j} \bar q_j\ , \\ \nonumber
&c_j := \frac{1}{\omega_j}(\omega_j^2 \zeta_j \zeta_{n-j} + z_j
z_{n-j}) = \frac{1}{2 \omega_j}(\bar p_{n-j}^2 - \bar p_j^2 +
\omega_j^2 \bar q_{n-j}^2 - \omega_j^2 \bar q_j^2) \ , \\ \nonumber 
&d_j := \frac{i}{\omega_j}(\omega_j^2 \zeta_j \zeta_{n-j} - z_j
z_{n-j}) = \frac{1}{\omega_j} (\bar p_j \bar p_{n-j} + \omega_j^2 \bar
q_j \bar q_{n-j}) \ , \\
\mbox{and if} \ n \ \mbox{is even} \nonumber \\
\nonumber
&a_{\frac{n}{2}} := i z_{\frac{n}{2}} \zeta_{\frac{n}{2}} = \frac{1}{2
\omega_{\frac{n}{2}}}(\bar p_{\frac{n}{2}}^2 + \omega_{\frac{n}{2}}^2
\bar q_{\frac{n}{2}}^2)\ . 
\end{align}
Note that these basis-elements are subject to the relation 
\begin{equation}
a_j^2 = b_j^2 + c_j^2 + d_j^2 
\end{equation}
\noindent and that $H_2$ can easily be expressed as 
\begin{equation}
H_2 = \sum_{1 \leq j \leq \frac{n}{2}} \omega_j a_j \ .
\end{equation}
\noindent Our definitions diagonalise the action of $S^*$:
\begin{equation}\label{Sopabcd}
S^*: a_j \mapsto a_j, \ a_{\frac{n}{2}} \mapsto a_{\frac{n}{2}}, \ b_j
\mapsto -  b_j, \  c_j \mapsto c_j, \ d_j \mapsto -d_j \ .
\end{equation}
\noindent The products $a_j a_k$, $a_{\frac{n}{2}} a_j$ and $b_j b_k$
are invariant under $S^*$ and $T^*$. The products $a_j b_k$ and
$a_{\frac{n}{2}} b_k$ are not invariant under $S^*$, although they are
under $T^*$.  It is left as an easy excercise for the reader to prove that
the only configuration for other terms to appear is $d_j
d_{\frac{n}{2} - j} - c_j c_{\frac{n}{2} - j}$.\\
\\
\noindent We summarize the results of this section in the following
theorem:
\begin{theorem}\label{eenvoud}
Let $H$ be the reduced periodic FPU Hamiltonian (\ref{gereduceerdeham}). There
is a fourth order normal form $\overline{H}$ of $H$ which is invariant
under $T^*$ and  $S^*$ \footnote{and $R^*$}. For this normal form we
have $\overline{H}_3 = 
0$, whereas $\overline{H}_4$ is a linear combination of the fourth
order terms $a_j a_k$, $b_j b_k$ $(1 \leq j , k < \frac{n}{2})$ and if
$n$ is even $a_{\frac{n}{2}} a_j$ $(1 \leq j \leq \frac{n}{2})$ and $d_j
d_{\frac{n}{2} - j} - c_j c_{\frac{n}{2}-j}$ $(1 \leq j \leq
\frac{n}{4})$. 
\end{theorem}

\section{Near-integrals or integrals of the truncated normal
form}\label{integrals} 
In the previous section we proved that the truncated fourth order
normal form of 
the periodic FPU Hamiltonian is subject to many restrictions, as
indicated in theorem \ref{eenvoud}. This enables us to point out some
integrals for the truncated
normal form. These are near-integrals of the periodic FPU chain:
quantities that are nearly conserved by the flow of the orginal chain
(\ref{gereduceerdeham}) for a
long time,
cf. \cite{verhulst}. \\ 
\indent In order to be able to compute these integrals, we first write down
the commutation relations between
the real basis-elements (\ref{abcd}). They are given by

\begin{equation}\label{commutatierelaties}
\{b_j, c_j\} = 2 d_j \ , \ \ \{b_j, d_j\} = - 2 c_j \ , \ \ \{c_j, d_j
\}= 2 b_j \ . 
\end{equation}
\noindent All the other Poisson brackets between basis-elements give
$0$. These relations lead to the following conclusions:

\subsection{The odd chain} 
\begin{corollary}\label{corr1}  If $n$ is odd, then the truncated normal form 
$H_2 + \overline{H}_4$ of the periodic FPU chain is Liouville
integrable with the quadratic 
integrals $a_j, 
b_j$ $(1 \leq j \leq \frac{n-1}{2})$.
\end{corollary}
\noindent {\bf Proof:} 
$H_2$ is a linear combination of the quadratics $a_j$ and $\overline{H}_4$ is a
linear combination of the fourth order terms 
$a_j a_k$ and $b_j b_k$. The $a_j$ and $b_k$ Poisson commute with all
these terms and with each other. $\hfill \square$
\\
\\
It is well-known (cf. \cite{arnold}), that the integrals of a
$2n-2$-dimensional 
Liouville integrable Hamiltonian system generally define
$n-1$-dimensional invariant tori. Let's see what these tori look like here and
analyse the integral map $F:
\mathbb{R}^{2n-2} \to \mathbb{R}^{n-1}$ that maps $(\bar q, \bar p)
\mapsto (a, b)$:
\begin{proposition}\label{imF}
\begin{equation}\label{imf}
\mbox{im}\ F = \{ (a,b) \in \mathbb{R}^{n-1} | a_j \geq 0, |b_j| \leq
a_j \} \ .
\end{equation}
For $(a, b) \in (\mbox{im}\ F)^{int} =  \{ (a,b)
\in \mathbb{R}^{n-1} | a_j > 0, |b_j| < a_j \}$, 
$F^{-1}(\{(a,b)\})$ is a smooth $n-1$-dimensional torus .

\end{proposition}
\noindent {\bf Proof:} Clearly, $\mbox{im}\ a_j = [0, \infty)$. The
level set of $a_j$ is, for $a_j > 0$, the cartesian product of
$\mathbb{R}^{2n-6}$ and a $3$-dimensional sphere in 
$\mathbb{R}^4$ with
radius $\sqrt{2 a_j}$. Let us consider $b_j$ restricted to the level
set of
$a_j$. To compute its critical points, we use the method
of Lagrange multipliers: $(\bar q, \bar p)$ is a critical point iff
there is a constant $\lambda$ such that $Da_j(\bar q, \bar p) =
\lambda Db_j(\bar q, \bar p)$, that is $(\omega_j \bar q_j, \omega_j
\bar q_{n-j}, \frac{1}{\omega_j} \bar p_j,\frac{1}{\omega_j} \bar
p_{n-j})$ $ = \lambda (- \bar p_{n-j}, \bar p_j, \bar q_{n-j}, - \bar
q_j)$. From this we infer that $\lambda = \pm 1$. For $\lambda = 1$,
we find $\bar p_{n-j} = - \omega_j \bar q_j, \ \bar p_j = \omega_j
\bar q_{n-j}$. In these points we have $b_j = a_j$. $\lambda = -1$
gives $\bar p_{n-j} =  \omega_j \bar q_j, \ \bar p_j = -\omega_j 
\bar q_{n-j}$, so $b_j = - a_j$. These are the extreme values of $b_j$
on the level set of $a_j$, giving (\ref{imf}). We also learn from this
that if $a_j >0$ 
and $|b_j| < a_j$, then $Da_j$ and $Db_j$ are independent. So if $(a,
b) \in (\mbox{im} F)^{int}$, then all $Da_j$ and $Db_k$ are independent
on $F^{-1}(\{(a, b)\})$. According to a theorem of Arnol'd (cf. \cite{arnold})
such a level set must be a torus. $\hfill \square$
\\
\\
\noindent In order to compute the flow on these tori, we make the explicit
transformation to action-angle coordinates $(\bar q, \bar p) \mapsto
(a, b, \phi, \psi)$ as follows. Let $\mbox{arg}: \mathbb{R}^2
-\{(0,0)\} \to \mathbb{R} / _{2 \pi \mathbb{Z}}$ be the argument
function, $\mbox{arg}: (r \cos \Phi, r \sin \Phi) \mapsto \Phi$. Then,
for $1 \leq j \leq \frac{n-1}{2}$, define
\begin{align}\label{actiehoek} \nonumber
&a_j := \frac{1}{2 \omega_j}
(\bar p_j^2 + \bar p_{n-j}^2 + \omega_j^2 \bar q_j^2 + \omega_j^2 \bar
q_{n-j}^2) \ , \\ 
&b_j := \bar p_j \bar
q_{n-j} - \bar p_{n-j} \bar q_j\ , \\ \nonumber
&\phi_j := \frac{1}{2} \mbox{arg} ( -\bar
p_{n-j} - \omega_j \bar q_j, \bar
p_j - \omega_j \bar q_{n-j} ) + \frac{1}{2}
\mbox{arg} (\bar 
p_{n-j} - \omega_j \bar q_j, \bar
p_j + \omega_j \bar q_{n-j} ) \ ,   \\ \nonumber
&\psi_j := \frac{1}{2} \mbox{arg} ( -\bar
p_{n-j} - \omega_j \bar q_j, \bar
p_j - \omega_j\bar q_{n-j} ) - \frac{1}{2}
\mbox{arg} (\bar 
p_{n-j} - \omega_j \bar q_j, \bar
p_j + \omega_j \bar q_{n-j} ) \ .
\end{align}

\noindent Note that these are well-defined as long as $(a, b) \in
(\mbox{im} F)^{int}$. With the formula $d\ \mbox{arg} (x,y) = \frac{x
dy - y dx}{x^2 + 
y^2}$, one can verify that the $(a, b, \phi, \psi)$ are canonical
coordinates: $\tilde 
\sigma = \sum_{j=1}^{n-1} d \bar q_j \wedge d \bar p_j = \sum_{1 \leq j
\leq \frac{n-1}{2}} d \phi_j \wedge d a_j + d \psi_j \wedge d b_j$. \\
\\
\noindent The truncated normal form $H_2 + \overline{H}_4$ is a function of
the actions $a_j, b_j$. Its induced equations of motion therefore read:
\begin{align}
\dot a_j = &\ \dot b_j = 0 \ , \\
\dot \phi_j = \omega_j + \frac{\partial \overline{H}_4}{\partial
a_j}(a, b) & \
,\ \ \dot \psi_j =  \frac{\partial \overline{H}_4}{\partial b_j} (a, b)
\ , \nonumber
\end{align}

\noindent which are very simple. In order to verify that that the
truncated normal form $H_2 + \overline{H}_4$ is nondegenerate, we
examine the frequency map $\omega$ which adds to each invariant torus
the frequencies of the flow on it:

\begin{equation} \nonumber
\omega: (a, b) \ \mapsto \left( \omega_1 + \frac{\partial
\overline{H}_4}{\partial a_1}(a, b), \ldots, \omega_{\frac{n-1}{2}} +
\frac{\partial \overline{H}_4}{\partial a_{\frac{n-1}{2}}}(a, b),
\frac{\partial \overline{H}_4}{\partial b_1}(a, b), \ldots,
\frac{\partial \overline{H}_4}{\partial b_{\frac{n-1}{2}}}(a, b) \right) \ .
\end{equation}

\noindent $\omega$ is a local diffeomorphism iff both the constant derivative
matrices $\frac{\partial^2 \overline{H}_4}{\partial a_j \partial a_k}$
and $\frac{\partial^2 \overline{H}_4}{\partial b_j \partial b_k}$ are
invertible. We will explicitly check this for the odd $\beta$-chain in
section \ref{betaketen}. \\ 
\\
\noindent  The situation is more difficult in the case of
\subsection{The even chain}
\begin{corollary}\label{corr2} If $n$ is even, then the truncated normal form $H_2
+ \overline{H}_4$ of the periodic FPU chain has 
the quadratic integrals $a_j$ $(1 \leq j \leq \frac{n}{2})$ and $b_j -
b_{\frac{n}{2} - j}$ $(1 \leq j < 
\frac{n}{4})$. 
\end{corollary}
\noindent {\bf Proof:}
$H_2$ is a linear combination of the quadratics $a_j$  $(1 \leq j \leq
\frac{n}{2})$, whereas $\overline{H}_4$ is a
linear combination of the 
fourth order terms $a_j a_k$ $(1 \leq j, k \leq \frac{n}{2})$, $b_j
b_k$ $(1 \leq j < \frac{n}{2})$  and
$d_j d_{\frac{n}{2}-j} - c_j c_{\frac{n}{2}-j}$ $(1 \leq j \leq
\frac{n}{4})$. The $a_j$ clearly commute with all these terms. So do
the terms $b_j - 
b_{\frac{n}{2}-j}$:  $\{b_j -
b_{\frac{n}{2}-j}, b_k\}$ $ = \{b_j -
b_{\frac{n}{2}-j}, a_k\}$  $ = \{b_j -
b_{\frac{n}{2}-j}, a_{\frac{n}{2}}\} = 0$. But one also verifies from
(\ref{commutatierelaties})  that 
 $\{ b_j -
b_{\frac{n}{2}-j}, c_j c_{\frac{n}{2}-j} - d_j d_{\frac{n}{2}-j}\} =$
$c_{\frac{n}{2}-j} \{b_j, c_j\} - c_j \{b_{\frac{n}{2}-j},
c_{\frac{n}{2}-j}\}$ $-d_{\frac{n}{2}-j} \{b_j, d_j\} +d_j
\{b_{\frac{n}{2}-j}, d_{\frac{n}{2}-j}\}$ $ = 2 c_{\frac{n}{2}-j} d_j
- 2 c_j d_{\frac{n}{2}-j}$ $+ 2 d_{\frac{n}{2}-j} c_j - 2 d_j
c_{\frac{n}{2}-j} = 
0$. $\hfill \square$  \\
\\
\noindent If $n$ is even, then the $n$-$1$-degrees of freedom
Hamiltonian $H_2 +
\overline{H}_4$ has at least $\frac{3n-4}{4}$ (if $4$ divides $n$)  or
$\frac{3n-2}{4}$ (if $4$ does not divide $n$) quadratic integrals. 
These are near-integrals for  
the original chain (\ref{gereduceerdeham}). We have not yet found a
complete system of integrals for the truncated normal form though. We
will do so for the even 
$\beta$-chain in section \ref{evenbetahoofdstuk}.

\section{The normal form of the $\beta$-chain}\label{betaketen}
In this section we present the explicit normal form of the periodic FPU
Hamiltonian in the case
that $H_3 = 0$, i.e. $\alpha = 0$ in
(\ref{hamfpu}). This chain, that has no cubic terms, is usually
referred to as the $\beta$-chain. A calculation of the normal form of
order $4$ is relatively easy in this case, because one does not have
to transform away $H_3$ first. The calculation is still tedious
though and that is why
we do not present it. The reader can find an example of a similar
computation in \cite{Rink}. The following theorem is a major
generalisation of the result in \cite{Rink}, which in turn is a
restatement -with a much more efficient proof-  of a theorem in the
PhD thesis of Sanders 
(\cite{sanders}).

\begin{theorem} If $\alpha = 0$, then in the periodic FPU chain one has
\begin{align}\label{normaalvormbeta} \nonumber
\overline{H}_4 = \frac{\beta}{n} \left\{ \sum_{0<k<l<\frac{n}{2}}
\frac{\omega_k \omega_l}{4} a_k a_l + \sum_{0<k<\frac{n}{2}}
\frac{\omega_k^2}{32}  (3 a_k^2 - b_k^2)  
+ \frac{1}{4} a_{\frac{n}{2}}^2 +
\frac{1}{2}a_{\frac{n}{2}}\sum_{0<k<\frac{n}{2}} \omega_k a_k
\right.  
\\
\left. +\frac{1}{8} \sum_{0 < k < \frac{n}{4}} \omega_{2k}^2
 (d_k d_{\frac{n}{2}-k} -  c_k c_{\frac{n}{2}-k})
+\frac{1}{16} (d^2_{\frac{n}{4}} - c^2_{\frac{n}{4}})
\right\} \ . 
\end{align}
\noindent In formula (\ref{normaalvormbeta}) it is understood that terms
with the subscript $\frac{n}{2}$ and $\frac{n}{4}$ only appear if $2$
respectively $4$ divides $n$.
\end{theorem}

\subsection{The odd $\beta$-chain}\label{onevenbeta}
In formula (\ref{normaalvormbeta}) we see again what was already
predicted in theorem \ref{eenvoud}, namely that $\overline{H}_4$ is a
linear combination of the terms $a_j a_k$ and $b_j b_k$ $(1 \leq j, k
\leq \frac{n-1}{2})$. According to corollary \ref{corr1} this normal form is
integrable, the $a_j$ and $b_j$ being the (quadratic) integrals. 
To check the nondegeneracy condition, we compute the second order derivative
matrices of $\overline{H_4}$ with respect to the action variables 
$a_j$ and $b_j$:

\begin{equation}\label{dhda}
\frac{\partial^2 \overline{H}_4}{\partial a_j \partial a_k} =
\frac{\beta}{16 \ n} \left( 
\begin{array}{cccc}
3 \omega_1^2 & 4 \omega_1 \omega_2 & \cdots &
4 \omega_1 \omega_{\frac{n-1}{2}} \\ 4 \omega_2 \omega_1 & 3
\omega_2^2 & \ldots &4 \omega_2 \omega_{\frac{n-1}{2}}  \\
\vdots & & \ddots & \vdots \\
4 \omega_{\frac{n-1}{2}} \omega_1  &  4 \omega_{\frac{n-1}{2}}
\omega_2  & \cdots &  
 3 \omega_{\frac{n-1}{2}}^2 
\end{array} \right) \ , 
\end{equation}

\begin{equation} 
\frac{\partial^2 \overline{H}_4}{\partial b_j \partial b_k} =
- \frac{\beta}{16 \ n} \left( 
\begin{array}{cccc} 
 \omega_1^2 & & & \\
 & \omega_2^2 & & \\
 & & \ddots & \\
 & & & \omega_{\frac{n-1}{2}}^2
\end{array} \right)
\end{equation}

\noindent $\frac{\partial^2 \overline{H}_4}{\partial b_j \partial
b_k}$ is clearly nondegenerate. But so is $\frac{\partial^2
\overline{H}_4}{\partial a_j \partial a_k}$. This can be proved by applying
elementary row and column operations to (\ref{dhda}), thus reducing it
to upperdiagonal form. This yields an expression for the
determinant that is unequal to $0$. We conclude that the reduced
periodic $\beta$-chain with an odd number of particles can, after a
near-identity transformation, be written as a perturbation of a
nondegenerate integrable Hamiltonian system. Therefore, the KAM theorem (cf. \cite{arnold})
applies:

\begin{theorem} \label{kam} If $n$ is odd, $\alpha = 0$ and $\beta
\neq 0$, then 
almost all low-energy solutions of the reduced periodic FPU chain
(\ref{gereduceerdeham}) are periodic or quasiperiodic and move on
invariant tori. In fact, the relative measure of all
these tori lying inside the
small ball $\{ 0 \leq  H \leq \varepsilon \}$, goes to $1$ as
$\varepsilon$ goes to $0$. 
\end{theorem}

\noindent  It should also be
possible to write down an expression for the normal form if $\alpha
\neq 0$. The nondegeneracy condition can be checked again then. But the
computation of this normal form is very nasty
- transforming away $H_3$ we obtain the transformed
$H_4'=H_4+\frac{1}{2}\{(\mbox{ad}_{H_2}|_{\mbox{im
ad}_{H_2}})^{-1}(H_3) , H_3\}$ which thereafter has to be normalised
to produce $\overline{H}_4$. The result is most likely that for almost
all $\alpha$ and $\beta$ the nondegeneracy condition holds and the KAM
theorem applies. Without computation
this is clear for $|\alpha| \ll |\beta|$, because in this situation
the coefficients of the normal form (\ref{normaalvormbeta}) change
only slightly and the invertible matrices form an open set in the
set of all matrices. 

\subsection{The even $\beta$-chain}\label{evenbetahoofdstuk}
It is a surprise that in
the normal form of the even $\beta$-chain no terms $b_j b_k \ (j\neq
k)$ arise, see formula (\ref{normaalvormbeta}). This
leads to the following remarkable conclusion:
\begin{corollary}\label{evenbeta}
If $n$ is even and $\alpha = 0, \ \beta \neq 0$, then the truncated
normal form $H_2 + \overline{H}_4$ of the reduced periodic FPU chain
(\ref{gereduceerdeham}) is Liouville integrable. The integrals are the
quadratics 
$a_j$ $(1 \leq j \leq \frac{n}{2})$,  $b_j -
b_{\frac{n}{2} - j}$ $(1 \leq j < \frac{n}{4})$ and $d_{\frac{n}{4}}$ (if
$n$ is a multiple of $4$) and the fourth order terms $\omega_k^2 b_k^2 +
\omega_{\frac{n}{2}-k}^2 b_{\frac{n}{2}-k}^2 + 4 \omega_{2k}^2
(c_k
c_{\frac{n}{2}-k} -  d_k
d_{\frac{n}{2}-k}) $ $(1 \leq j < \frac{n}{4})$.
\end{corollary} 

\noindent {\bf Proof:} This follows from simply computing all the
Poisson brackets,
using (\ref{commutatierelaties}) and the fact that the Poisson
brackets form a derivation. $\hfill \square$\\
\\
\noindent Only the $a_j, b_j - b_{\frac{n}{2}-j}$ and
$d_{\frac{n}{4}}$ induce a $2 \pi$-periodic flow and can  therefore be
seen as actions after some symplectic action-angle transformation.  It
is an open 
problem to construct the remaining action variables. Thereafter one
could differentiate $\overline{H}_4$ with respect to them and verify the
KAM nondegeneracy condition. \\
\\
One exceptional case is easier: the $\beta$-problem with $4$
particles. Its truncated 
normal form reads:
\begin{equation}
H_2 + \overline{H}_4 = \sqrt{2} a_1 + 2 a_2 + \frac{\beta}{4} \left( \frac{1}{8} a_1^2 +
\frac{1}{4} a_2^2 + \frac{\sqrt{2}}{2} a_1 a_2 + \frac{1}{8} d_1^2 \right) \ ,
\end{equation}
which has the commuting integrals $a_1, a_2$ and $d_1$. The frequency map is  
\begin{equation}
\omega: (a_1, a_2, d_1) \mapsto (\sqrt{2} + \frac{\beta}{16} a_1 +
\frac{\sqrt{2}\beta}{8} a_2, 2 + \frac{\beta}{8} a_2 + \frac{\sqrt{2}\beta}{8} a_1,
\frac{\beta}{16} d_1 ) \ .
\end{equation}
$\omega$ is nondegenerate, since 
\begin{equation} 
\frac{\partial \omega}{\partial (a_1, a_2, d_1)} = \frac{\beta}{4} \left(
\begin{array}{ccc} \frac{1}{4} & \frac{\sqrt{2}}{2}  & 0 \\
\frac{\sqrt{2}}{2} & \frac{1}{2} & 0 \\
0 & 0 & \frac{1}{4} \end{array}
\right) 
\end{equation}
\noindent is invertible. So a similar theorem as \ref{kam} holds for
the $\beta$-chain with $4$ particles. \\
\\
\noindent It is unclear what happens for the even chain if $\alpha
\neq 0$. The truncated normal form might not be integrable. On
the other hand we already found about $\frac{3n}{4}$ integrals. And in
\cite{Rink} it was already shown that the normal forms of the 
$\alpha$-$\beta$-chain with up to $6$ particles are Liouville integrable.

\section{Discussion}
The lesson that we can learn from this analysis is that the characteristic
features of the Fermi-Pasta-Ulam chain, such as quasiperiodicity and
nonergodicity, are not just a property shared by all
low-energy solutions of resonant Hamiltonian systems. On the contrary: the
periodic FPU chain is a rather special system possessing particular
symmetries and eigenvalues. These cause or may cause nondegenerate
integrability of the Birkhoff-Gustavson normal form of the chain,
which in turn implies that the KAM theorem (cf. \cite{arnold}) is applicable. Still, some questions
remain unanswered:\\
\\
\noindent {\bf 1.} From corollary \ref{corr1} we know that the
truncated normal form of the odd FPU chain is integrable. In section
\ref{onevenbeta} we checked a nondegeneracy condition for the odd
$\beta$-chain and were able to apply the KAM theorem. Can 
 the truncated normal form of the odd chain explicitly be computed
also if $\alpha 
\neq 0$? Is it really nondegenerate, as
we are tempted to assume? \\
\\
\noindent {\bf 2.} What is the reason that the truncated normal form
of the even $\beta$-chain is 
integrable as we know from corollary \ref{corr2}? Is there some hidden
symmetry-like property of the FPU 
chain that prevents terms $b_j b_k$ $(j \neq k)$ from appearing in the
truncated normal form, thus causing the integrability?\\
\\
\noindent {\bf 3.} Is it possible to explicitly construct action-angle
coordinates for 
the truncated normal form of the even $\beta$-problem, globally or locally, and
verify the KAM nondegeneracy condition?\\
\\
\noindent {\bf 4.}  What about the even $\alpha$-$\beta$ chain? As
indicated in corollary \ref{corr2} its truncated normal form has a lot
of conserved quantities. But is it also
really Liouville integrable? If yes, then there is a big chance for
the KAM theorem to 
work. And otherwise:
can we find a counterexample of an even $\alpha$-$\beta$ chain with many
ergodic orbits of low energy?\\ 
\\
\noindent Where the second question is of a rather philosophical
nature, the other three involve tough computations. Answers might be
given in a subsequent paper.

\newpage 
\appendix
\section{Proof of theorem
\ref{hoofdstellingeigenwaarden}} 
This appendix is based on notes of Frits Beukers. Its main intention
is to prove theorem \ref{hoofdstellingeigenwaarden}. Some algebra is
used that might be uncommon to the reader, but fortunately the
conclusions of theorem \ref{hoofdstellingeigenwaarden} and theorem
\ref{eenvoud} are easily
understood. 

\subsection{Sums of roots of unity}
We are interested in solving the resonance equation $\nu(\Theta, \theta) = 0$,
that is we want to find vanishing sums of the eigenvalues $\pm i
\omega_j= \pm 2 i\sin(\frac{j\pi}{n})$. A study of these sums is possible
if we first 
consider sums of roots of unity.\\ 
\indent Fix $N\in\mathbb{N}$. We study the equation
$\zeta_1+\zeta_2+\cdots+\zeta_N=0$ 
in the unknown roots of unity $\zeta_i$. The solutions will be
determined 
modulo permutation of the terms and multiplication by a common root of 
unity. We also assume that there are {\it no vanishing subsums}, that
is 
$\sum_{i\in I}\zeta_i\ne0 \ {\rm for\ all}\ I\subset\{1,\ldots,N\}, 
|I|<N$. We first state our basic tool. Let $K$ be
a field generated over $\mathbb{Q}$ by roots of unity. Let
$p^k$ be a prime power and let $\zeta:=e^{2\pi i/p^k}$.
Suppose $\zeta\not\in K$ and 
$\zeta^p\in K$. 

\begin{proposition}\label{minimaal} The minimal polynomial of $\zeta$
over the field $K$ 
is given by $X^p-\zeta^p$ if $k\ge2$ and $X^{p-1}+X^{p-2}+\cdots+X+1$
if $k=1$.
\end{proposition}

\noindent For the proof of this proposition we refer to
\cite{Waerden}, \S 60-61.\\ 
\indent To return to our problem let us choose $M\in\mathbb{N}$
minimal so that 
$(\zeta_i/\zeta_j)^M=1$ for all $i,j=1,2,\cdots,N$. Since we can  
multiply every
term of our relation with $\zeta_1^{-1}$ and put $\zeta_i=\zeta_i/\zeta_1$
we may as well assume that all $\zeta_i$ are $M$-th roots of unity.
Let $p^k$ be a
primary factor of $M$. Set $M'=M/p$ and  write 
$\zeta_i=\tilde{\zeta_i}\zeta^{n_i}$
where $\tilde{\zeta_i}\in K:= \mathbb{Q}(e^{2\pi i/M'})$ and
$n_i\in\{0,1,2,\ldots, 
p-1\}$. Then, according to proposition \ref{minimaal}, the minimal 
polynomial of $\zeta$ over $K$ is $X^p-\zeta^p$ if $k\ge2$ and
$X^{p-1}+X^{p-2}+\cdots+X+1$ if $k=1$. \\
\indent We now rewrite our relation in the following form
$$\sum_{s=0}^{p-1}\sum_{n_i=s}
\tilde{\zeta}_i \zeta^s=0\eqno{(R)} \ .$$

\noindent If $k\ge2$ the minimal polynomial of $\zeta$ over $K$ is
$X^p-\zeta^p$. 
In particular this means that there exist no non-trivial $K$-linear
relations 
between $1,\zeta,\ldots,\zeta^{p-1}$. So the relation (R) implies that
all coefficients are zero, hence
$\sum_{n_i=s}\tilde{\zeta_i}\zeta^{s}=0$
for all $s=0,1,2,\ldots,p-1$. By the minimal choice of $M$, at least
two of the exponents $n_i,n_j$ should be different. Hence the
assumption $k\ge2$ leads automatically to vanishing subsums.\\
\indent Let us now assume $k=1$. Then the minimal polynomial of
$\zeta$ over $K$ is 
$X^{p-1}+X^{p-2}+\cdots+X+1$. This means that any $K$-linear relation
between 
$1,\zeta,\ldots,\zeta^{p-1}$ must have all of its coefficients equal.
Hence, (R) implies that all sums
$$\sum_{n_i=s}\tilde{\zeta_i}\eqno{\rm(P)}$$
have the same value $\sigma$. Since we do not want vanishing subsums
we necessarily have $\sigma\ne0$. This in its turn implies that each of the
summations contains at least one term and so 
$p\le N$. This puts a bound on our search range. 

\subsection{Explicit computations}
In this section we compute vanishing sums of roots of unity having
no vanishing subsums. It should be noted that the solutions are given
modulo permutation of terms and multiplication by a common root of unity.\\
\indent For each of the specific values of $N$ we shall be considering, we
denote by $M$ the smallest number such that $(\zeta_i/\zeta_j)^M=1$
for $i,j$. From the previous section we know that $M$ is square free
and that $p\le N$ for all prime divisors of $M$. Furthermore, we
also note that if $M$ divides $6$, then it is easy to see that the
only possible relations without vanishing subsums
are $1-1=0$ and $1+\delta+\delta^2=0$ where $\delta=e^{2\pi i/3}$.
So we shall assume that there is a prime $\ge5$ dividing 
$M$. By $N\ge p\ge5$ the first interesting case to be considered is $N=5$.

\begin{itemize}

\item[$N=5$.] We have $5|M$. 
Then (P) partitions our sum in precisely five parts,
each with equal sum. Hence $1+\eta+\eta^2+\eta^3+\eta^4=0$ where
$\eta=e^{2\pi i/5}$. 

\item[$N=6$.] Then $p\le5$, hence $5|M$. Then (P) 
partitions our sum in four parts of length $1$ and one with length $2$.
Hence we see that $-\delta-\delta^2+\eta+\eta^2+\eta^3+\eta^4=0$ is the
solution. 

\item[$N=7$.] Then $p\le7$. If $7|M$ then necessarily,
$1+\epsilon+\epsilon^2+\epsilon^3+\epsilon^4+\epsilon^5+\epsilon^6=0$
where $\epsilon=e^{2\pi i/7}$. 

Suppose $5$ is the largest prime dividing $M$. Then (P) gives a partitioning
in $31111$ or $22111$. The first gives rise to solutions with zero subsums.
The second gives rise to the solutions $(-\delta-\delta^2)(1+\eta)+
\eta^2+\eta^3+\eta^4=0$ and $(-\delta-\delta^2)(1+\eta^2)+
\eta+\eta^3+\eta^4=0$. 

\item[$N=8$.] Then $p\le7$. If $7|M$ then (P) implies that we have
a partitioning $2111111$ and $-\delta-\delta^2+
\epsilon+\epsilon^2+\epsilon^3+\epsilon^4+\epsilon^5+\epsilon^6=0$.

Suppose $5$ is the largest prime dividing $M$. Then (P) gives a
partitioning $41111$, $32111$ or $22211$. The first two give rise only
to vanishing subsums. The last solution gives rise to
$(-\delta-\delta^2)(1+\eta^i+\eta^j)+\eta^k+\eta^l=0$ where
$\{i,j,k,l\}=\{1,2,3,4\}$. 
\end{itemize}

\subsection{Sums of the $i \omega_j$}\label{eigenwaardensom}
We are interested in vanishing sums of the eigenvalues $\pm i \omega_j
= \pm 2
i \sin(\frac{j \pi}{n})$. So we look for all solutions of 
$\zeta_1+\cdots+\zeta_N=0$ such that together with each $\zeta_i$, minus
its complex conjugate $-\zeta_i^{-1}$ also occurs. Since we shall only
be interested in sums of $3$ or $4$ eigenvalues $i \omega_j$, we
restrict ourselves to $N=6,8$. We include sums with vanishing 
subsums, except vanishing subsums of the form $\zeta-\zeta=0$, since
these give rise to vanishing subsums of $i \omega_j$'s. 
So all vanishing subsums of roots of unity must have length at least three.  

\begin{itemize}
\item[$N=6$.] To bring our relation without zero subsums in the
desired form, we have to multiply it by $\pm i$ and we derive 
$$2 i \sin(\pi/6) + 2 i \sin(\pi/10) - 2 i \sin(3\pi/10)=0 \ .$$
Now we look at relations with vanishing subsums. There can only be two
vanishing subsums of length three. Hence 
$(\zeta_1+\zeta_2)(1+\delta+\delta^2)=0$ with $\zeta_1,\zeta_2$
arbitrary. It is necessary and sufficient to assume that $\zeta_1\zeta_2=-1$.
This means
$(\zeta - \zeta^{-1})(1+\delta+\delta^2)$ where $\zeta$ is
arbitrary. Hence, 
$$2 i \sin(\pi r)+ 2 i \sin(\pi(r+2/3))+ 2 i \sin(\pi(r+4/3))=0 \ ,$$
where $r$ is an arbitrary rational number.

\item[$N=8$.] Let us first see what we get from our relations without
zero subsums. We find 

$$2 i \sin(\pi/6)+ 2 i \sin(3\pi/14)-2 i \sin(\pi/14)- 2 i \sin(5\pi/14)=0$$
$$2 i \sin(\pi/6)+ 2 i \sin(13\pi/30)- 2 i \sin(7\pi/30)- 2 i \sin(3\pi/10)=0$$
$$2 i \sin(\pi/6)+ 2 i \sin(\pi/30) - 2 i \sin(11\pi/30)+ 2 i
\sin(\pi/10)=0 \ .$$ 

Any relation with vanishing subsums must have subsums both of length 4,
or subsums of lengths $3$ and $5$. The first case cannot occur, but the
second yields $\zeta_1(1+\delta+\delta^2)+\zeta_2(1+\eta+\cdots+\eta^4)=0$.
Both $\zeta_1,\zeta_2$ must be purely imaginary and have opposite
sign. So we can 
take $\zeta_1= -\zeta_2=i$, hence 
$$2 i \sin(\pi/2) - 2 i \sin(\pi/6)+ 2 i \sin(\pi/10) - 2i
\sin(3\pi/10)=0$$  
\end{itemize}

\subsection{Proof of theorem \ref{hoofdstellingeigenwaarden}}
We indicate how theorem \ref{hoofdstellingeigenwaarden} can
be proved using the previous paragraphs. \\ 
\indent From the first relation in section \ref{eigenwaardensom} we
infer that $i \omega_{\frac{n}{6}} + i\omega_{\frac{n}{10}} - i
\omega_{\frac{3n}{10}} = 0$ if $n$ is a multiple of $30$. So
multi-indices $\Theta, \theta$ can be found such that $|\Theta| +
|\theta| = 3$ and $\pm \nu(\Theta, \theta) = i\omega_{\frac{n}{6}} + i
\omega_{\frac{n}{10}} - i \omega_{\frac{3n}{10}} = 0$. But for this
$\Theta$ and $\theta$, we must have that $\mu(\Theta, \theta) = \pm
\frac{n}{6} \pm  \frac{n}{10} \pm \frac{3n}{10}$ of which one easily
verifies that it is unequal to $0$ modulo $n$. \\ 
\indent One finds the same result for the other third order relation
of the previous section. The verification is not hard, but needs more
bookkeeping because of the appearance of the arbitrary rational. The
conclusion is that for all multi-indices $\Theta, \theta$ with
$|\Theta| + |\theta| = 3$ and $\nu(\Theta, \theta) = 0$, we have that
$\mu(\Theta, \theta) \neq 0$ mod $n$. This proves the first part of
theorem \ref{hoofdstellingeigenwaarden}, which actually states that
$P_3 \cap \mbox{ker ad}_{H_2}$ is too small to have a nontrivial
intersection with $\mbox{ker}(T^* - \mbox{Id})$. \\ 
\indent The proof of the second part of
theorem \ref{hoofdstellingeigenwaarden} is not harder. For $|\Theta| +
|\theta| = 4$, there are a number of trivial solutions to the equation
$\nu(\Theta, \theta) = 0$, namely those of the form $i \omega_j -
i \omega_j + i \omega_k -i \omega_k = 0$. These give rise to the
$\Theta$ and $\theta$ mentioned in theorem
\ref{hoofdstellingeigenwaarden}. All the other solutions to $\nu = 0$
must be of the form mentioned in section \ref{eigenwaardensom} under
the item `$N=8$'. From the first relation we see for instance that $i
\omega_{\frac{n}{6}} + i \omega_{\frac{3n}{14}} - i \omega_{\frac{n}{14}} -
i\omega_{\frac{5n}{14}} = 0$ if $n$ is a 
multiple of $42$. So multi-indices $\Theta, \theta$ with $|\Theta| +
|\theta| = 4$ can be found such that $\pm \nu(\Theta, \theta) = i
\omega_{\frac{n}{6}} + i \omega_{\frac{3n}{14}} - i \omega_{\frac{n}{14}} -
i\omega_{\frac{5n}{14}} = 0$. But for these multi-indices, one must
have that $\mu(\Theta,\theta) = \pm \frac{n}{6} \pm \frac{3n}{14} \pm
\frac{n}{14} \pm \frac{5n}{14} \neq 0$ mod $n$. One finds the same
conclusion for the other relations under the item `$N=8$'. This poves
the second part of theorem \ref{hoofdstellingeigenwaarden}.
\\ 
\\

\end{document}